# Accelerated Discovery of Materials with Extreme Work Functions through Uncertainty-Aware Multi-Fidelity Screening

**Authors:** Jun Meng,[1,*] Ryan Jacobs,[1,*] Rehan Kapadia,[2] John Booske[3]

[1] Department of Materials Science and Engineering, University of Wisconsin-Madison, Madison, WI, USA

[2] Department of Electrical and Computer Engineering, University of Southern California, Los Angeles, USA, 90089

[3] Department of Electrical and Computer Engineering, University of Wisconsin-Madison, Madison, WI, USA

*Corresponding author e-mail: jmeng43@wisc.edu, rjacobs3@wisc.edu

## Abstract

Work function plays a pivotal role in technologies ranging from energy conversion and electronics to catalysis. In this work, we integrated machine learning (ML) with multi-fidelity screening to develop a data-driven framework for accelerating the discovery of materials with extreme work functions. We augmented a previously published Random Forest (RF) model for work function to include prediction uncertainty calibration and domain of applicability assessment to enhance prediction robustness. By combining the augmented RF model with universal ML interatomic potential simulations and targeted *ab initio* calculations, we screened ≈5.5 million compounds from the GNoME and Alexandria databases. This workflow identified 209 surfaces with extreme low work functions below 2.0 eV and 227 surfaces with extreme high work functions above 6.0 eV, corresponding to 136 and 172 unique materials, respectively. The resulting candidates revealed trends consistent with established chemical principles, including the tendency of alkali- and alkaline-earth-terminated surfaces to exhibit low work functions. While it also uncovered less conventional motifs: lanthanide-rich surface terminations were strongly associated with extremely low work functions, whereas surfaces containing metalloids or phosphorus at the top layer were correlated with exceptionally high work functions. This work demonstrates a scalable strategy that leverages ML models and multi-fidelity computational efforts to accelerate the discovery of materials with extreme work functions for advanced electronic, energy-conversion, and catalytic applications.

## 1. Introduction

The work function $\Phi$ is a fundamental electronic property of a material's surface, and quantifies the minimum energy required to remove an electron from the the bulk Fermi level to the vacuum level outside the material. Work function therefore governs electron exchange at interfaces (both material-vacuum and material-material) and plays a pivotal role across technologies such as solid-state electronics, vacuum electronics, thermionic energy conversion, photovoltaics, photoelectrochemical cells, and catalysis.[1–9] Materials with low work functions ($\Phi < 2.5\,eV$) efficiently emit electrons, making them essential for thermionic and field emitters, thermionic conversion devices, and for contacts to emerging semiconductor materials.[2,4,6] Since the emission current density is exponentially sensitive to $\Phi$, determined by the Richardson–Laue-Dushman equation for thermionic emission, or Fowler-Nordheim equation for field emission, even reductions of as little as 100-200 meV in $\Phi$ can lead to orders of magnitude increases in emission current at a fixed temperature or allow for lower temperature operation at a fixed current, enhancing both energy efficiency and device longevity. Beyond electron emission, materials with low work functions can act as catalysts for reduction reactions. As strong electron donors, materials like perovskite oxides and MXenes exhibit low work functions, facilitate efficient electron transfer, and boost catalytic performance in various reactions, particularly in electrocatalysis.[9–16] Conversely, high work function materials (> 6.0 eV) serve as effective electron sinks, making them critical to create efficient hole-transport layers (HTLs) or electron-blocking layers in organic light-emitting diodes (OLEDs), organic photovoltaics (OPVs), and perovskite solar cells.[17,18] By matching the material's Fermi level to the frontier orbitals of the active organic or perovskite layer, high-$\Phi$ electrodes minimize the energy barrier for charge injection or extraction, thereby improving device efficiency and reducing operating voltage.

Work function is highly sensitive to surface-specific characteristics, making its measurement and prediction inherently complex. First, it exhibits pronounced crystallographic anisotropy whereby different crystal facet orientations (e.g., (100), (110), (111)) of the same material can present varying atomic arrangements and surface dipoles, leading to a range of work function values. The impact of surface orientation on the work function can also vary widely based on material type. For example, metals tend to only show 0.1-0.3 (typically < 0.5 eV) variations in work function between different facet orientations,[19,20] while more ionically bonded solids such as perovskite oxides may show eV-scale differences in work function.[1,12] Surface atoms often deviate from ideal bulk terminations through relaxation or reconstruction, altering surface dipoles and significantly influencing the work function, which, for oxides, can again have an eV-scale effect.[15,21] Additionally, the presence of adsorbates, even at sub-monolayer coverage, can drastically change the work function. For example, the commercialized thermionic dispenser cathodes used for various high power vacuum electronic applications use W metal as a base material, with an average work function of 4.5 eV, but a sub-monolayer coverage of Ba-O adsorbates reduces the measured effective work function to $\approx$2 eV. These complexities pose significant challenges for both experimental characterization and computational prediction. Experimental techniques such as Ultraviolet Photoelectron Spectroscopy (UPS), and Kelvin Probe Force Microscopy (KPFM) are standard tools, but their results can be difficult to interpret for micro- or nano-scopically heterogeneous surfaces such as polycrystalline samples, single crystals with multiple exposed facets or terminations, or environmental conditions like high temperature or specific gas ambient that are expected to alter the work function. On the computational side, Density Functional Theory (DFT) using a slab model is the standard approach for calculating work functions. However, the accuracy of these calculations is dependent on numerous factors, including the choice of exchange-correlation functional (e.g. Perdew-Burke-Ernzerhof (PBE) vs. Heyd-Scuseria-Ernzerhof (HSE)), the thickness of the slab, and the size of the vacuum region. Despite these complexities, there is significant physical understanding of what qualitatively affects a material's work function. For example, the presence of surface adsorbed electropositive species like alkali and alkaline metals typically lowers the work function, while electronegative species such as halogens typically increase the work function. However, there are exceptions to these basic trends, and detailed approaches for understanding and predicting work function on a per-material and even per-surface facet basis are still needed.[22]

The past decade has witnessed the acceleration of materials discovery by the advent of high-throughput computational screening. Enabled by the confluence of powerful computing resources, automated workflow software, maturation of data science and machine learning, and large, open-access databases of DFT-calculated properties like the Materials Project,[23,24] C2DB,[25] AFLOW,[26] and OQMD,[27] researchers can now systematically explore vast chemical spaces. This data-driven approach has accelerated the pace of discovery in diverse application spaces, from superconductors and thermoelectrics to catalysts and battery materials. This revolution has extended to the study of materials with extreme work functions.[13,14,28,29]

Early DFT studies mapped the landscape of work functions for various material classes, leading to the successful identification of promising low work function material candidates. Some of the first studies in this space attempting to find novel classes of extreme work function materials came simultaneously in 2016 from Jacobs *et al*.[12] and Hansmann *et al*.[9] These initial studies only examined a small number of potential perovskite oxide materials (order ~20 different compositions), but found that materials like $SrVO_3$ and $BaZrO_3$ had promising low work functions of about 2 eV and 0.82 eV, respectively.[9,12] As a next step, Ma *et al*. used DFT method and fast-to-calculate electronic structure descriptors that serve as a proxy for work function, screened over 2900 perovskite oxides in search of stable, conductive, low-work-function materials, and identified $BaMoO_3$ and $SrNb_{0.75}Co_{0.25}O_3$ with work functions of 1.1 eV and 1.5 eV, respectively.[11] Moving beyond purely DFT-based screening, a landmark contribution came from Schindler *et al*., who created the largest database of calculated work functions to date (58,332 surfaces from 3,716 materials) and used it to train the first random forest (RF) machine learning (ML) model for rapid work function prediction.[4] This model enabled rapid discovery of 34 ultra-low and 56 ultra-high work function surfaces, and successfully identified the (100)-Ba-O termination of $BaMoO_3$ with work function of 1.25 eV, consistent with a previous DFT study.[11] To further improve the predictive accuracy, Hsu *et al.* introduced a Force-Informed, Relaxed Equivariance Graph Neural Network (FIRE-GNN) that enables accurate and rapid prediction of materials surface properties, including work function and cleavage energy across vast chemical spaces.[30] By integrating surface-normal symmetry breaking and machine learning interatomic potential (MLIP)-derived force information, this architecture achieves a twofold reduction in mean absolute error compared to the prior RF model.

While these advancements have significantly increased the speed of work function prediction, assessing the reliability of model predictions remains critical for autonomous materials discovery. In particular, the uncertainty estimates from an RF model reflect the spread of predictions across individual decision trees in the ensemble, but does not necessarily correspond to the true prediction error ($\Phi_{ML} - \Phi_{DFT}$). As a result, the raw uncertainty estimate from an RF model can provide misleading confidence in the predicted work function values. In this work, we augmented the RF model built upon the previous study[4] with calibration of the uncertainty estimates and domain of applicability assessment. Uncertainty calibration produced error bars that are statistically consistent with the observed errors. Domain of applicability classification separated in-domain (ID) predictions from out-of-domain (OD) predictions, allowing candidates that have less trustworthy predictions to be filtered out. Leveraging this augmented RF model, in combination with a graph-network model for bandgap prediction,[31] high-throughput surface relaxation using the universal machine learning interatomic potential (U-MLIP), and targeted DFT calculations, we screened approximately 5.5 million compounds from the Graph Networks for Materials Exploration (GNoME)[32] and Alexandria materials databases.[33] This multi-fidelity screening workflow enabled reliable and accelerated discovery of 219 surfaces with low work functions (below 2.0 eV) and 227 surfaces with high work functions (above 6.0 eV), corresponding to 136 and 172 unique materials, respectively, which are of potential interest for various applications requiring extreme work function materials.

## 2. Methods

### 2.1 Random Forest model augmentation with uncertainty calibration and domain of applicability assessment

The RF ML model was trained and analyzed using the MAterials Simulation Toolkit for Machine learning (MAST-ML),[34] which is an open-source software package designed with a particular focus on regression models for predicting materials properties, including state-of-the-art approaches for uncertainty calibration and domain of applicability assessment. Following the work by Schindler *et al.*,[4] the RF model was trained on the dataset comprising 58,332 DFT-calculated work functions of unrelaxed surfaces. The use of unrelaxed surfaces was necessary to keep the amount of DFT computational time tractable. Relaxation of the terminating surface may result in atomic structures with corresponding work functions that differ significantly from those obtained in the unrelaxed variant of the surface, and we address this issue more below in Sec. 2.2. The features were chosen from the previous work,[4] including surface atomic features such as electronegativity, inverse atomic radius, first ionization energy, and Mendeleev number of elements present at the top and second layers; and surface geometric features such as packing fraction per layer, distances between atomic layers, and angles associated with surface atom arrangement. The performance of the RF model was assessed by random 5-fold cross-validation (CV), leave-out-material CV, and leave-out-extreme WF CV to evaluate the generalization of the model across different data regimes. Feature importances are analyzed by using the Shapley Additive ExPlanations (SHAP) approach.[35]

The RF model comprises an ensemble of decision trees, where each tree is trained on a different bootstrapped subset of the training data. The predicted work function ($\Phi_{ML}$) is the mean of the individual tree predictions, while the standard deviation among these predictions serves as the raw uncertainty estimate $\widehat{\sigma}_{raw}$. However, these raw uncertainty estimates are not guaranteed to match the observed true error, defined here by the residual between the ML prediction and the DFT-calculated value, $\Phi_{ML} - \Phi_{DFT}$. To obtain uncertainty estimates that accurately reflect these observed errors, we calibrated the RF model's $\widehat{\sigma}_{raw}$ using the calibrated bootstrap method proposed by Palmer *et al.*,[36] which has demonstrated high effectiveness in uncertainty (i.e. error bar) calibration for RF models across many datasets in the materials science and engineering field.[37] The calibrated uncertainty estimate $\widehat{\sigma}_{cal}$, was obtained by linearly rescaling $\widehat{\sigma}_{raw}$ through a log-likelihood optimization procedure such that the resulting $z$-scores ($(\Phi_{ML} - \Phi_{DFT})/\widehat{\sigma}_{cal}$), follow a standard normal distribution as closely as possible.

In addition, it is essential to assess the domain of applicability as a *guardrail* to flag predictions that are sufficiently dissimilar from the training data, which often exhibit high prediction errors, poorly calibrated uncertainty ($\widehat{\sigma}_{cal}$), or both due to extrapolation. Here, we use the recently developed method by Schultz *et al.*,[38] which quantifies how *dissimilar* a new data point is from the training data distribution. The method works by first fitting a Kernel Density Estimation (KDE) model to the feature vectors of the entire training dataset. The KDE provides a non-parametric representation of the probability density function of the training data in the high-dimensional feature space. When a new candidate material is evaluated, its feature vector is assessed using this KDE model. The dissimilarity is calculated as the negative log-likelihood of the test point feature vector. A high dissimilarity score suggests that the point lies in a region of low probability density, far from the bulk of the training data, and therefore may be considered out-of-domain (OD). This metric provides a powerful tool to automatically flag predictions that are less reliable due to the poor extrapolation of RF model, allowing them to be filtered out before committing to more expensive computational or experimental validation.

## 2.2 Multi-Fidelity High-throughput Screening Workflow

We screened approximately 5.5 million compounds from the GNoME and Alexandria databases using a multi-fidelity screening framework. Candidate materials were first filtered for non-radioactive, thermodynamically stable metallic compounds using database-derived stability metrics and a graph neural network band gap model.[31] The resulting materials were then screened using the uncertainty-aware RF model for rapid work function prediction, followed by surface relaxation with the U-MLIP M3GNet,[39] and targeted DFT calculations for high-fidelity work function evaluation. An overview of the screening approach and number of materials passing each criterion is shown in Figure 1. The GNoME database consists of 384,872 stable compounds on the convex hull as calculated by DFT, and the Alexandria database consists of 5,068,744 DFT-relaxed compounds in the 3D inorganic portion of the database. We first applied three preliminary screening criteria to the compound pool: (1) excluding compounds containing radioactive elements (Ac, Th, Pa, U, Np, Pu, Am, Cm, Bk, Cf, Es, Fm, Md, No, Lr, Rf, Db, Sg, Bh, Hs, Mt, Ds, Rg, Cn, Nh, Fl, Mc, Lv, Ts, Og, Tc, Pm, Po, At, Rn, Fr, Ra); (2) retaining only thermodynamically stable entries with energy above the convex hull $E_{\mathrm{hull}} \leq 50$ meV/atom, and (3) restricting the dataset to metallic materials only as the RF model was trained on metallic materials. The work function for metals is unambiguously the energy difference of vacuum level to Fermi level, while for semiconductors and insulators, the minimum energy required for electron emission may range from the ionization energy (i.e. the energy difference of vacuum level to valence band maximum) to the electron affinity (i.e., the energy difference of vacuum level to conduction band minimum). Accurate band-edge placement for non-metallic materials requires higher-fidelity DFT (e.g., with hybrid functionals), careful k-point sampling, and stricter electrostatic controls. In addition, semiconductors and insulators are more prone to exhibit complex surface reconstructions, which can lead to a drastically different work function from that obtained with simple surface cleaving, resulting in computational predictions that don't represent the work function that would be observed in the actual material. Mixing metals with semiconductors or insulators therefore may result in misleading rankings. For practical reasons, we restricted the research interest to metallic compounds in this study, and emphasize that more thorough screening and analysis of non-metals is a worthy topic for future work. To restrict the search to metallic materials, we first excluded compounds with nonzero DFT-PBE band gaps reported in the GNoME and Alexandria databases. To further reduce false metallic classifications arising from the well-known band gap underestimation of PBE, we applied a graph neural network model trained on band gap data spanning multiple fidelity levels, including DFT-PBE, DFT-HSE, and experimental measurements.[31] The graph-network model was used to predict high-fidelity HSE-equivalent band gaps, and compounds with predicted $E_{gap}^{\mathrm{HSE}} > 0$ eV were excluded from further consideration.

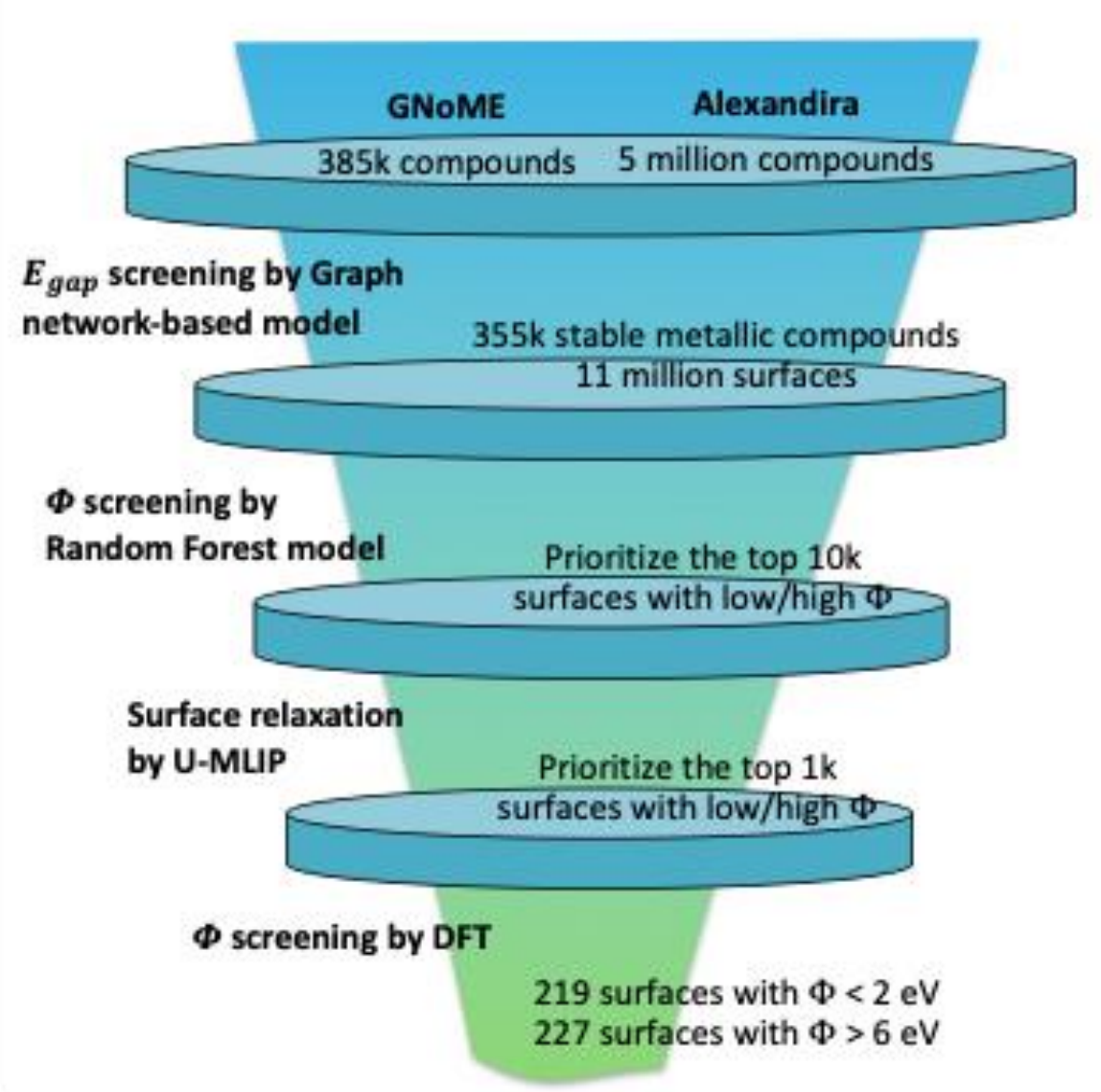

*Figure 1 Schematic of high-throughput screening workflow for materials with extreme work functions.*

After these initial screening steps, we retained 143,469 bulk compounds from the GNoME database and 211,353 bulk compounds from the Alexandria database. For each bulk material, we enumerated symmetry-distinct, low-index surfaces (i.e. (100), (001), (010), (110), (101), (111)) using the WFRFModel toolkit from Schindler *et al.*,[4] to generate slab terminations and the 15 near-surface descriptors (elemental and geometry descriptors of the top three layers). These bulk compounds yielded approximately 6.5 million and 4.2 million low-index surfaces from GNoME and Alexandria database, respectively. The augmented ML model was then used to predict the work function $\Phi$, the calibrated uncertainty (error bar) $\hat{\sigma}_{cal}$, and the domain of applicability dissimilarity score $d$ for each surface. Based on analysis of results in Sec. 3.3, only predictions with $d \leq 0.97$ were considered in-domain (ID) and retained.

### 2.3 Universal Machine Learning Interatomic Potential (U-MLIP) and Density Functional Theory (DFT) calculations

After removing surfaces that were flagged as out-of-domain (OD), we prioritized approximately 10,000 surfaces each for the low work function regime below 2.2 eV and the high work function regime above 5.8 eV. These surfaces were constructed with a total vacuum thickness of 40 Å along the surface normal direction, distributed symmetrically as 20 Å on the top side and 20 Å on the bottom side of the slab. This symmetric vacuum spacing minimizes spurious interactions between periodic images and provides a sufficient vacuum plateau. These surfaces were relaxed using M3GNet,[39] which is trained on a diverse dataset consisting of DFT calculations for 62,783 compounds from Materials Project and provides near-DFT accuracy but with orders of magnitude lower computational cost. The M3GNet relaxations were performed by allowing only the top surface region to relax, while keeping the bottom part of the slab fixed to approximate the bulk substrate. Specifically, atoms were sorted by their fractional coordinate along the surface normal direction, and only the top 1/3 atoms were allowed to relax, while the bottom 2/3 were constrained during optimization. The relaxation was considered converged when the maximum force on each atom was below 0.05 eV/Å.

After M3GNet relaxation, DFT calculations were performed using the projector augmented wave method[40] as implemented in the Vienna Ab initio Simulation Package (VASP).[41,42] The PBE generalized gradient approximation exchange-correlation functional[43] was applied for a single-shot static self-consistent field (SCF) electronic structure calculation. The kinetic energy cutoff for the plane wave basis was set to 520 eV. Gamma-centered k-point meshes were generated based on the slab lattice vectors with the k-points sampling density of 0.04 $Å^{-1}$ in the Brillouin zone. To account for asymmetric slab geometries and avoid artificial electric fields arising from periodic boundary conditions, dipole corrections were applied along the surface normal direction. Electronic convergence was achieved with an energy tolerance of $10^{-6}$ eV/cell. Gaussian smearing with a small smearing parameter of 0.05 eV was applied for the SCF calculation. The work function was calculated as $\Phi = V_{\text{vac}} - E_F$ from the converged electrostatic potential in the vacuum region and the Fermi level, respectively. The vacuum potential was extracted from the plateau region of the planar-averaged electrostatic potential along the surface normal direction.

## 3. Results and Discussion

### 3.1 Performance of the augmented RF model and interpretation of features

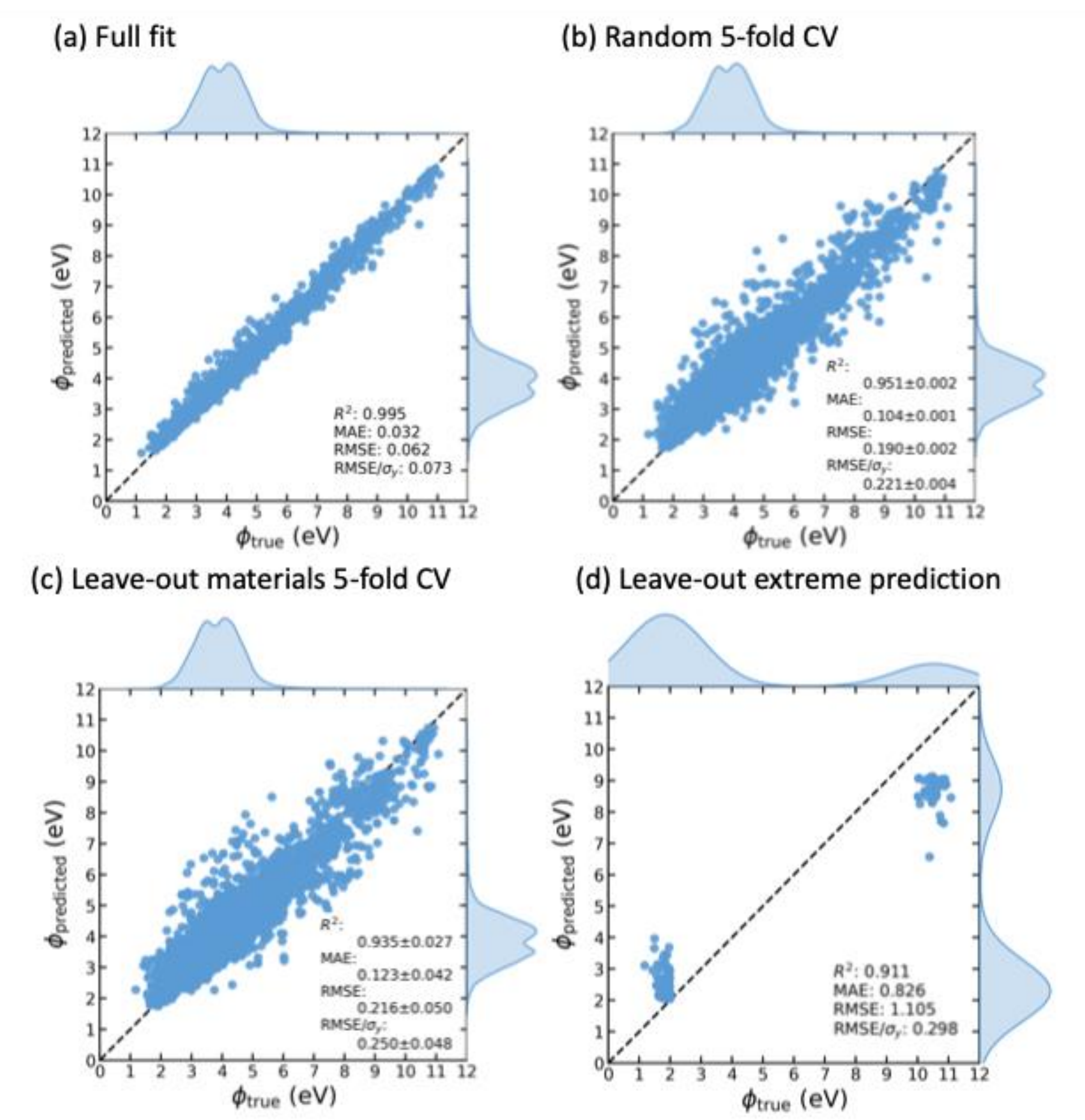

*Figure 2 Performance evaluation and feature importance analysis of the machine learning model for work function prediction. Parity plots for (a) the full fit, (b) random 5-fold cross-validation, (c) leave-out material group 5-fold cross-validation, (d) leave-out extreme work functions prediction.*

To ensure the reliability and robustness of the augmented RF ML model before large-scale screening, we conducted a comprehensive evaluation encompassing baseline performance benchmarking, cross-validation (CV) under extrapolative regimes, and assessment of uncertainty and domain metrics. Using the original dataset of 58,332 DFT-calculated work functions, the full augmented model demonstrates excellent agreement between predicted and ground truth DFT-calculated work functions on the full dataset (Figure 2a), achieving a coefficient of determination $R^2$ of 0.995 and a low mean absolute error (MAE) of 0.032 eV, consistent with the reported value of 0.03 in the previous study.[4] Robustness was further verified through cross-validation strategies. The random 5-fold cross-validation (CV) yields only a modest degradation of accuracy with $R^2$ of 0.951 and MAE of 0.104 eV (Figure 2b), demonstrating strong generalization when the training and test sets share similar chemistries. In the leave-materials-out CV (Figure 2c), surfaces from the same bulk material were excluded in each fold to evaluate the model's ability to generalize to new surfaces from unseen materials. This CV test is the most indicative of how we envision most researchers would employ this model and how we do so in our screening, where one is interested in predicting the work function of a totally new materials not in the training dataset. The MAE increased only slightly to 0.12 eV compared to the random 5-fold CV, indicating strong generalization to unseen materials. However, the model exhibits limited extrapolation capability when tested on extreme values outside the training distribution. In Figure 2d, when trained on surfaces with work functions between 2 to 10 eV and tested on those below 2 eV or above 10 eV, the model exhibited a negative $R^2$ and an MAE of 1.87 eV on the test dataset, showing the difficulty of extrapolation beyond the training domain. This extrapolation difficulty is due to the reason that RF models' predictions are based on averages of training set targets within terminal tree leaves. As a result, predictions for values outside the training data are often pulled back toward the training distribution, i.e. the tendency to underpredict high work functions and overpredict low work functions. This limitation motivates the use of domain of applicability assessment to support trustworthy high-throughput screening, as discussed below. These results suggest that when materials with a target work function regime are present in the training data, the RF model can effectively identify new materials and surface terminations with similar work functions by interpolation within the learned chemical space. This capability is valuable for discovering alternative candidate materials near known low or high work function regimes.

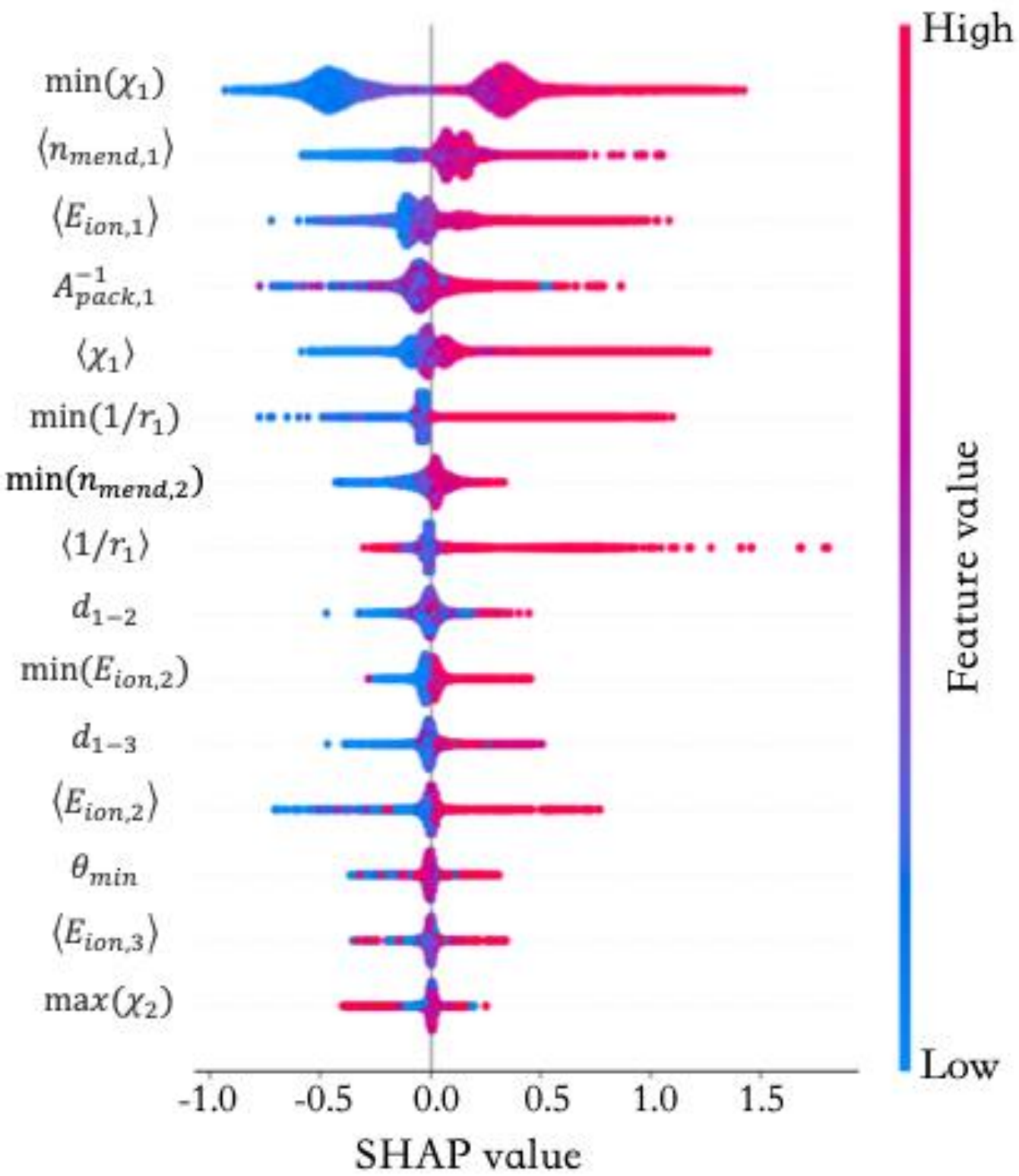

*Figure 3. SHAP summary plot illustrating feature importance and impact on model predictions. Features are ranked vertically from top to bottom based on their mean absolute SHAP value, which represents the overall magnitude of the feature's impact. Each point represents an individual data sample. The color indicates the actual feature value (red for high, blue for low), while the horizontal axis displays the SHAP value. Positive values increase the predicted work function, while negative values decrease it.*

To investigate the factors governing the prediction accuracy of the model, we analyzed feature contributions using SHAP (SHapley Additive exPlanations), which quantify how much each feature shifts an individual prediction. Figure 3 ranks features by their overall importance, measured as the mean absolute SHAP value across all predictions, and shows the contribution of each feature. The minimum electronegativity among all elements at the top layer $\min(\chi_1)$, turns out to be the most influential feature. Surfaces with high values of $\min(\chi_1)$ (red dots) overwhelmingly contribute to positive SHAP values, increasing the prediction of work function. Whereas low $\min(\chi_1)$ surfaces (blue points) contribute to negative SHAP values, and lower the prediction. This analysis shows that the prediction of work function is highly sensitive to the minimum electronegativity at the top layer. This aligns exactly with the well-known physical expectations that elements with low electronegativity often have low work functions. Mendeleev number $\langle n_{mend,\ 1}\rangle$ and the first ionization energy $\langle E_{ion,1}\rangle$ are the following high-rank features, indicating that surfaces enriched with elements possessing deeper valence states or stronger electron binding tend to exhibit higher work functions. Collectively, the top three features account for 44% of the total RF feature importance, while $\min(\chi_1)$ alone contributes 26%, demonstrating that the prediction of work function is predominantly governed by elemental electronic descriptors associated with the terminating surface. The inverse atomic radius descriptors $\min(1/r_1)$ and $< \frac{1}{r_1} >$ also contribute positively, indicating that small-size ions (such as highly charged cations) will likely elevate the work function. Beyond these leading variables, geometric descriptors such as the inverse packing fraction (number of atoms per unit cell area) $A_{pack,1}^{-1}$, and interlayer distances $d_{1-2}$ and $d_{2-3}$ exhibit moderate positive trends, showing that less densely packed surfaces and smaller interlayer spacings increase the prediction. These effects are secondary to those of the dominant electronic descriptors but reflect subtle structural correlations captured by the RF model.

### 3.2 ML model uncertainty recalibration

As discussed in Sec. 2.1, the RF model's raw uncertainty $\widehat{\sigma}_{raw}$ is estimated from the standard deviation of the individual tree predictions, which is not guaranteed to be consistent with the observed true error ($e = \Phi_{ML} - \Phi_{DFT}$). Therefore, we calibrated the raw uncertainty to better correlate with the true error statistically using the calibrated bootstrap method.[36] Figure 4a compares the distribution of the $z$-scores ($z$-scores = $e/\widehat{\sigma}$) for the leave-out test data before calibration using $\widehat{\sigma}_{raw}$ (grey), and after calibration using $\widehat{\sigma}_{cal}$ (blue). For a well-calibrated model, the $z$-score should follow a standard normal distribution, where it has the mean of 0 and standard deviation of 1. The mean remains close to zero both before and after calibration, with values of −0.001 and 0.0207, respectively, indicating little systematic bias in the RF model's prediction, i.e. $\Phi_{ML}$. Before calibration, however, the standard deviation is 0.676, substantially below the ideal value of 1. This indicates that the $\widehat{\sigma}_{raw}$ are systematically larger than the true errors. After calibration, the standard deviation increases to 0.999, demonstrating that the $\widehat{\sigma}_{cal}$ reflect the true errors statistically more accurately.

Figure 4b provides a complementary assessment of calibration quality by plotting the $z$-score distributions as a function of binned reduced uncertainty $\widehat{\sigma}/\sigma_y$, which is normalized by the standard deviation of the dataset. The plotted symbols indicate the mean $z$-score within each bin, and the error bars indicate the corresponding standard deviation. This visualization highlights how this calibration improves the uncertainty estimates across the entire range of uncertainty magnitudes. Notably, calibration compresses the range of reduced uncertainty values from approximately 0 - 3 before calibration to roughly 0 - 2 after calibration. The broad range of $\widehat{\sigma}_{raw}/\sigma_y$, together with the deviation of the $z$-score distributions from the ideal standard-normal behavior at $\frac{\widehat{\sigma}_{raw}}{\sigma_y} > 2.2$, indicates that the $\widehat{\sigma}_{raw}$ were larger compared with the actual scale of work function variation in the dataset. After calibration, the $\widehat{\sigma}_{cal}/\sigma_y$ are compressed into a smaller range, and most bins show $z$-score distributions that are close to the ideal standard normal distribution. This demonstrates that the calibration improves the statistical consistency between the $\hat{\sigma}_{cal}$ and the true errors. Figure 4c evaluates the overall quality of the uncertainty estimates by quantifying the discrepancy (i.e. miscalibration area) between the cumulative distribution function (CDF) of the $z$-scores and that of a standard normal distribution. Smaller value indicate better agreement of $z$-scores distribution with the standard normal distribution. Across the full range of reduced uncertainty, the calibrated $\hat{\sigma}_{cal}$ generally exhibits lower miscalibration areas, i.e. better agreement with the normal distribution, than the uncalibrated $\hat{\sigma}_{raw}$. The average miscalibration area decreases from approximately 0.08 before calibration to 0.053 after calibration, showing that the calibrated uncertainties more accurately capture the distribution of true errors.

Figure 4d evaluates the correlation between the true error and the reduced uncertainty, as well as the impact of calibration on improving this relationship. The plot shows reduced root mean square residuals (i.e. true errors) versus binned reduced uncertainty estimates, providing a direct measure of how well predicted uncertainties track true errors. Ideally, perfect correlation would yield a slope of 1 and an intercept of 0. The uncalibrated model yields a slope of 0.73, substantially below the ideal value of 1, indicating systematic overestimation of true errors. After calibration, the slope improves markedly to 1.09, much closer to the ideal, showing that the magnitude of the uncertainty estimates now closely matches the true errors. It is worth noting that overestimation of error bar magnitudes before calibration is a common trait of RF models, at least when applied to materials property prediction.[37] Overall, these results show that the calibration significantly improves the quantitative accuracy of the uncertainty estimates, bringing the calibrated uncertainties much closer to agreement with the true errors.

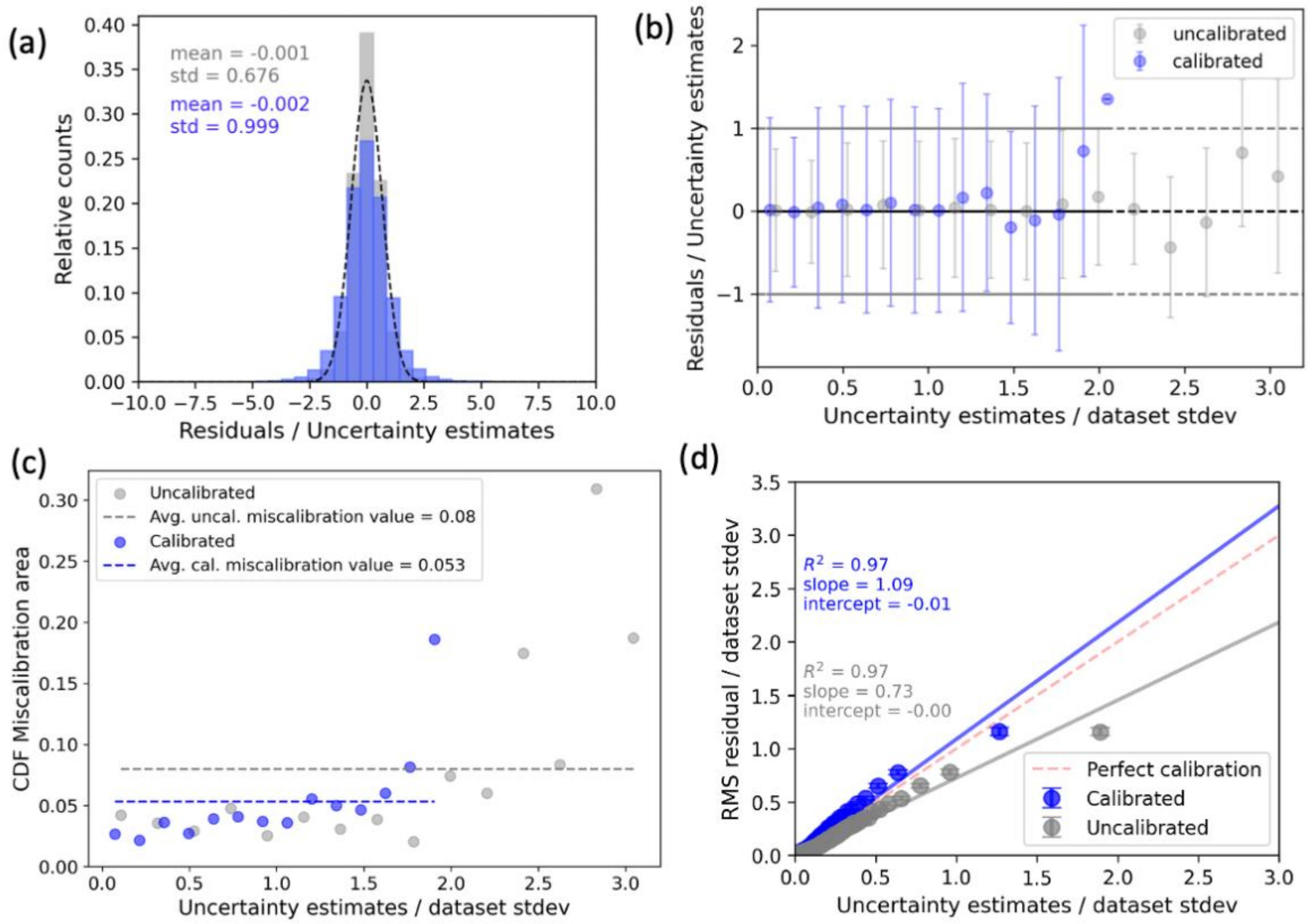

*Figure 4. Calibration of RF model uncertainty estimates. For all plots, the grey and blue data correspond to data before and after*

*calibration, respectively. (a) Distribution of $z$-scores before and after calibration, where the $z$-score is the observed true error divided by the RF model's uncertainty estimate. A well-calibrated model should yield a standard normal distribution with mean of 0 and standard deviation of 1. (b) $z$-score distributions as a function of reduced uncertainty, normalized by the standard deviation of the dataset. Points show the mean $z$ -score in each bin, and error bars show the corresponding standard deviation. (c) Miscalibration area as a function of reduced uncertainty, measuring the deviation between the observed $z$ -score cumulative distribution function (CDF) and an ideal standard normal CDF; smaller values indicate better calibration. (d) Reduced RMS residuals as a function of reduced uncertainty. The dashed red line indicates perfect calibration. Calibration brings the RF uncertainty estimates into closer agreement with the true errors. Each bin contains 1006 data points, with 174 bins in total.*

### 3.3 ML model domain

Next, we evaluated the domain of applicability (DoA) for each prediction using the method developed by Schultz *et al.*,[38] which is implemented in the MAST-ML framework. In this approach, the kernel density estimate (KDE) is used to determine the distance of the candidate from the training dataset. The resulting KDE distance, $d$, is then related to the reduced root-mean-square-error $E^{RMSE}/_{\sigma_y}$, which reflects the prediction accuracy, and the miscalibration area $E^{area}$, which reflects the predicted error bar accuracy. Candidates that lie farther from the training distribution generally exhibit larger $E^{RMSE}/\sigma_y$ and $E^{area}$, indicating reduced reliability in both the predicted work functions and their associated uncertainty estimates.[37,38] In this approach, Schultz *et al.*[38] determine the cutoff by comparing candidate dissimilarity thresholds against a chosen ground truth of in-domain (ID) vs. out-of-domain (OD) labels, and selecting the threshold that maximizes classification quality, i.e., F1 score. Note that there is no universal ground truth to determine ID vs. OD (e.g., Schultz et al. use a $E^{RMSE}/\sigma_y$ value of 1.0), and it is up to the user to decide on a value that corresponds to a tolerable error. Here, our goal is to focus only on highly reliable predictions, so we apply a threshold of $d$=0.97 at which the $E^{RMSE}/\sigma_y$ reaches approximately 0.5. This choice is motivated by the observation that $E^{RMSE}/\sigma_y$ generally increases with distance from the training data, and values below about 0.5 correspond to very good performance.[37] Thus, using $E^{RMSE}/\sigma_y \approx 0.5$ as the operative cutoff provides a practical criterion to maximize our chances of success in finding new extreme work function materials.

Figure 5a-b show binned reduced RMSE and miscalibration area as functions of KDE distance ($d$). Both plots reveal that as KDE distance increases, the model's accuracy declines (higher reduced RMSE) and error bar reliability decreases (higher miscalibration area). Using $d$ = 0.97 as the cutoff, ID points are denoted in green and OD in red. Figure 5c provides a confusion matrix, which shows how well the KDE-distance cutoff, $d$=0.97, separates reliable ID predictions from OD predictions. The diagonal entries represent correctly classified points, while the off-diagonal entries represent misclassifications. The model correctly classifies the correct domain about 95.4% (84% and 9.4% for ID and OD correct classification, respectively) of the time, with a false positive rate (i.e., the undesirable classification where the model predicts a point as ID when really it is OD) of only 0.64%. Figure 5d further illustrates the correlation between KDE distance and both reduced RMSE and miscalibration area, confirming that larger KDE distance values correspond to poorer predictive accuracy and less reliable uncertainty estimates.

Finally, Figure 5e summarizes domain classification of the whole training dataset. The histogram distributions of the ID and OD classifications for the training dataset show that most of the training data are classified as ID, confirming that the KDE-based domain criterion does not excessively exclude points within the original training distribution. OD points constitute only a small fraction of the dataset and are distributed across the whole range. However, the surfaces at the extreme high end of the distribution ($\Phi$ > 7 eV) are all flagged as OD, suggesting reduced reliability in these regions. The domain assessment analysis provides a powerful, data-driven justification as a practical guardrail. Any new prediction flagged as OD is automatically deprioritized for computationally expensive follow-up studies. This strategy enables an efficient and prioritized discovery workflow by focusing resources on the most reliable predictions.

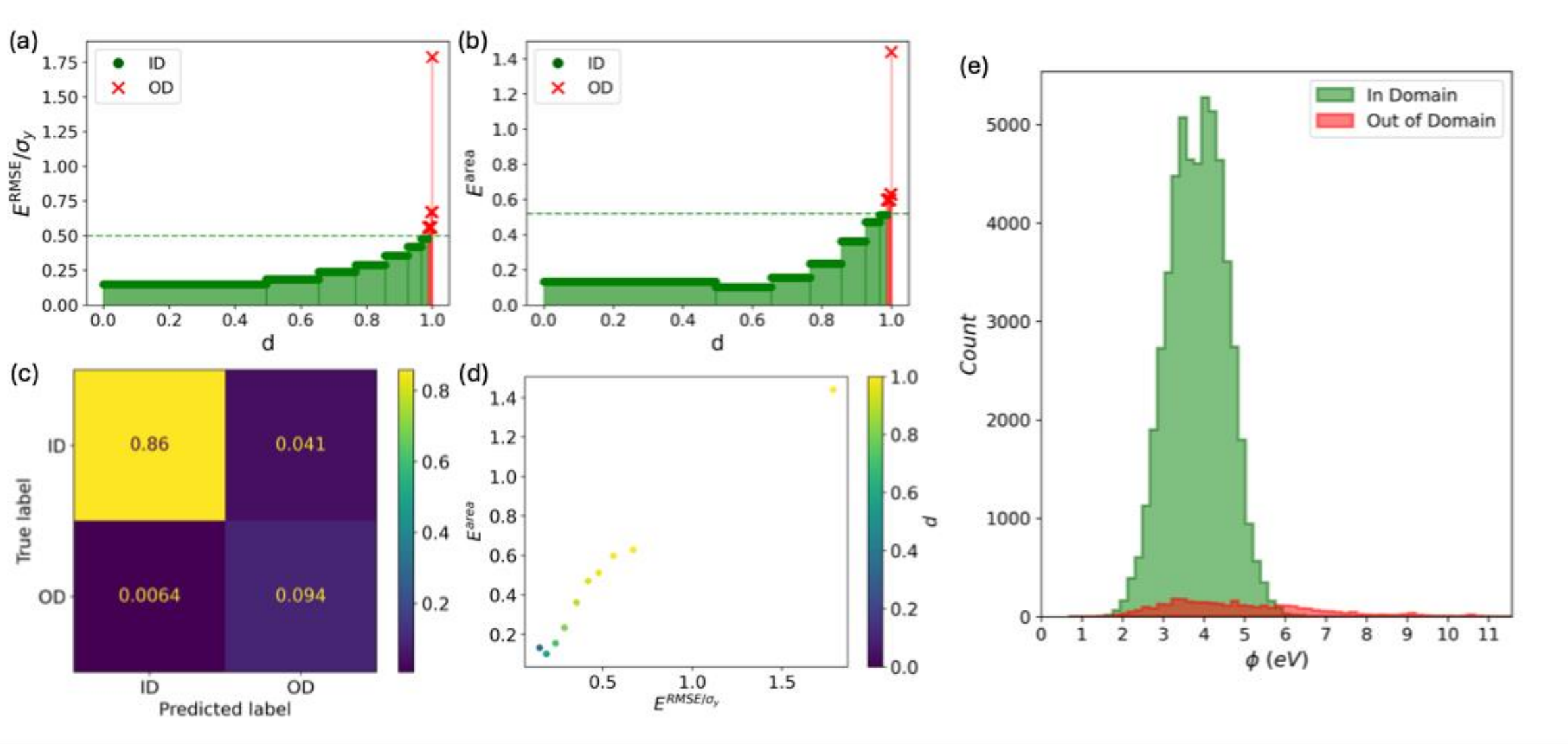

*Figure 5. Domain of Applicability assessment. (a) Plot of binned reduced RMSE vs. KDE feature distance. (b) Plot of binned reduced error bar miscalibration areas vs. KDE feature distance. In (a) and (b), the green points denote in domain (ID), and the red point denote out of domain (OD). The green dashed line is the ground truth threshold between in and out of domain. (c) Confusion matrix for classification of in- vs. out-of-domain, expressed as a fraction of the total data. (d) Plot of reduced error bar miscalibration area vs. reduced RMSE with color denoting KDE feature distance. (e) Domain of applicability of data in the training dataset.[4]*

### 3.4 Multi-fidelity Screening on Metallic Compounds

We screened two open-source materials databases: GNoME, comprising approximately ≈385k DFT-calculated stable crystals (using the PBE functional), and Alexandria, containing 5.07 million DFT-relaxed 3D inorganic structures. As described in Sec. 2.2, we applied the following filters: (i) excluding compounds containing radioactive elements; (ii) retaining only thermodynamically stable compounds; and (iii) restricting the dataset to metallic compounds. These filters yielded a refined search space of ≈355k metallic candidates, a ≈44% reduction from the initial set. It is worth noting that the training dataset and the screened GNoME and Alexandria compounds cover the same set of chemical elements, spanning approximately 80 elements across the periodic table, including alkali metals, alkaline earth metals, transition metals, metalloids, chalcogens, halogens, and lanthanides, except noble gases and actinides.

For each compound, low-index surfaces ((100), (001), (010), (110), (101), (111)) were enumerated using the WFRFModel toolkit,[4] generating ≈11 million unique surfaces from 355k bulk materials. Each surface was characterized with the 15 descriptors previously identified as most relevant to work function. The augmented RF model was then applied to predict three key quantities for each surface: the work function, the calibrated uncertainty, and the domain-of-applicability dissimilarity distance score $d$. Surfaces with $d$ above 0.97 were flagged as OD and excluded. This filtering retained ≈10.5 million surfaces, with up to 95% classified as ID, demonstrating the model's broad applicability.

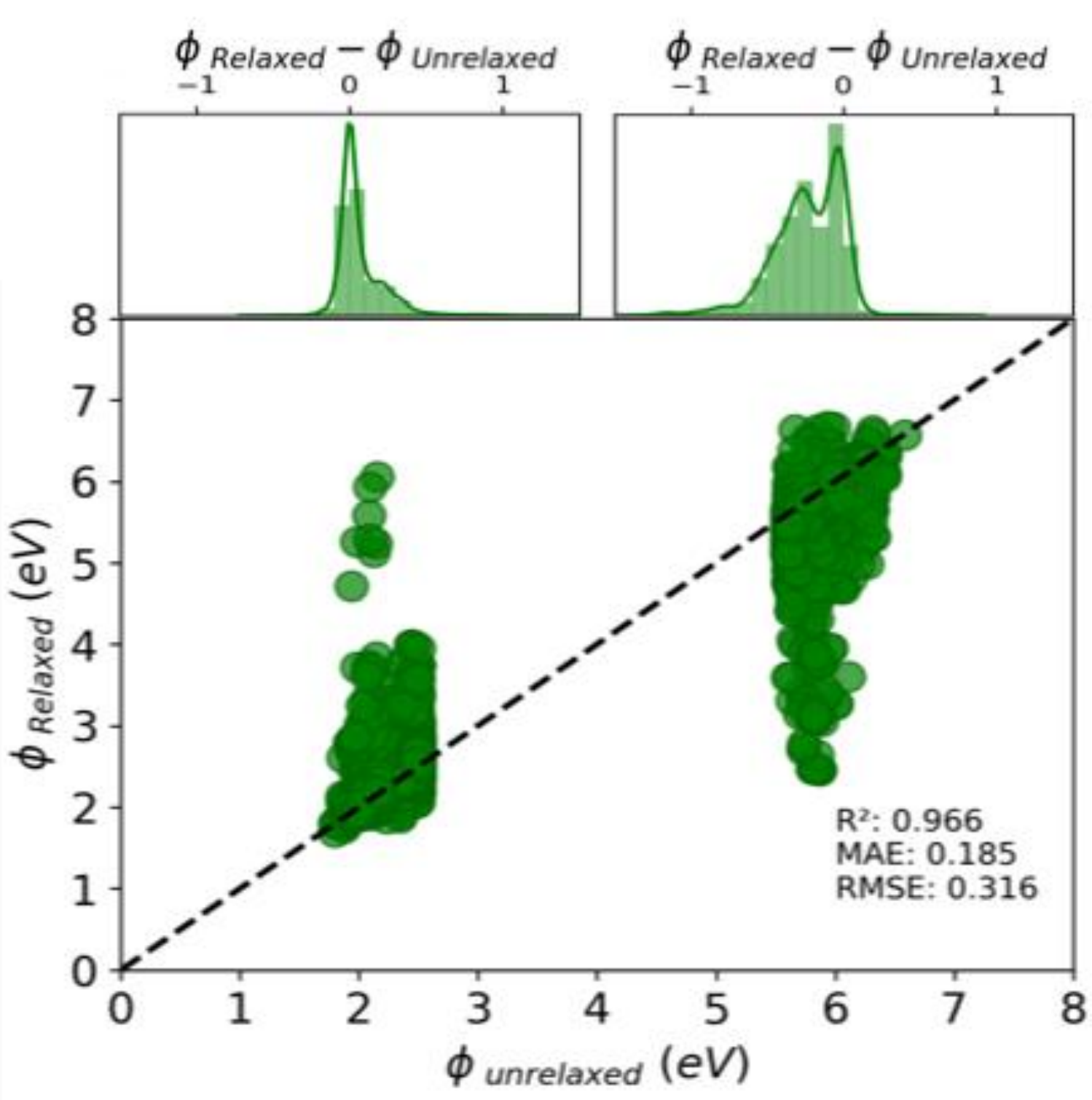

*Figure 6. Parity plot comparing ML-predicted work functions on relaxed surfaces vs. unrelaxed surfaces, with residual distribution for extreme low and high work functions.*

We next focused on materials with extreme predicted work functions by selecting the ≈20k highest- and lowest-ranked surfaces for subsequent relaxation calculations, ensuring that subsequent calculations remained computationally tractable. This screened subset comprised 6.3k surfaces with $\Phi_{ML}$ below 2.5 eV and 4.1k surfaces with $\Phi_{ML}$ above 5.8 eV from GNoME dataset, together with 5k surfaces with $\Phi_{ML}$ below 2.2 eV and 5k surfaces with $\Phi_{ML}$ above 5.6 eV from the Alexandria database. We then performed surface relaxation using the M3GNet, serving as a computationally efficient approach for capturing realistic surface geometries. We then re-predicted work functions for relaxed surfaces. From the GNoME database, 3.4k out of 6.3k remained with $\Phi_{ML}^{relaxed} \leq 2.5$ eV (46% reduction), while only 377 surfaces retained with $\Phi_{ML}^{relaxed} \geq 5.8$ eV (92% reduction). From the Alexandria database, 3.5k surfaces retained with $\Phi_{ML}^{relaxed} \leq 2.5$ eV (30% reduction), and 1.5k surfaces retained with $\Phi_{ML}^{relaxed} \geq 5.6$ eV (69% reduction).

In Figure 6, we show the difference between predicted work functions of unrelaxed and relaxed surfaces, where surface relaxation substantially modifies the predicted work functions and, therefore, cannot be neglected when screening for extreme values. For surfaces with low unrelaxed work functions, relaxation typically produces ±0.2 eV shifts, despite the fact that a subset of surfaces exhibits larger increases. In contrast, surfaces with high unrelaxed work functions experience a downward shift of 0.5-0.6 eV upon relaxation, with a few cases showing modest increases (<0.2 eV). While the absolute magnitudes of these changes differ between the low and high work function regimes, the corresponding relative changes are both on the order of 10% of the actual work function values. The magnitude of these findings is in good agreement with the previous work, where they found that surface relaxation can raise low work function values by 0.2 eV and lower high work function values by 0.6 eV.[4] Overall, these results highlight that surface relaxation is a critical component of the screening workflow, as it can significantly alter candidate materials that pass through thresholds.

Finally, we conducted the static self-consistent DFT calculations on the retained ≈1.5k relaxed surfaces using the PBE functional. First, however, we further narrowed the candidate pool by applying more restrictive thresholds to the relaxed surfaces, focusing only on 383 surfaces with $\Phi_{ML}^{relaxed} \leq 2.2$ eV and 377 surfaces with $\Phi_{ML}^{relaxed} \geq 5.8$ eV from the GNoME database, and 322 surfaces with $\Phi_{ML}^{relaxed} \leq 2.0$ eV and 438 surfaces with $\Phi_{ML}^{relaxed} \geq 5.8$ eV from the Alexandria database. These DFT-PBE calculations yielded 209 surfaces with $\Phi_{DFT} < 2.0$ eV and 227 surfaces with $\Phi_{DFT} > 6.0$ eV. The top 10 candidates with the lowest/highest work functions are presented in Table 1 along with their associated surface index and terminating elements. The retained dataset constitutes a set of high-confidence, novel material candidates for in-depth investigation.

**Table 1.** DFT-PBE calculation work functions for metallic material surfaces with extreme work functions.

| **Materials with low work function** | | | | |
|---|---|---|---|---|
| Composition | Surface Index | Surface Element | DFT Work Function(eV) | Source |
| $Rb_2CdIn_4Au_3$ | 001 | Rb | 1.28 | Alexandria |
| $La_6NdS_8$ | 110 | La, S | 1.50 | Alexandria |
| $La_6PrS_8$ | 110 | La, S | 1.50 | Alexandria |
| $RbIn_6Pt$ | 001 | Rb | 1.58 | Alexandria |
| $NaPr_3S_4$ | 010 | Pr, S | 1.66 | GNoME |
| $Pr_6SmS_8$ | 110 | Pr, S | 1.66 | Alexandria |
| $La_4Pr_3S_8$ | 100 | La, Pr, S | 1.67 | Alexandria |
| $RbIn_6Cu$ | 001 | Rb | 1.67 | Alexandria |
| $Pr_6NdS_8$ | 110 | Pr, S | 1.68 | Alexandria |
| $La_5AsS_4$ | 001 | La, S | 1.68 | Alexandria |
| **Materials with high work function** | | | | |
| Composition | Surface Index | Surface Element | DFT Work Function(eV) | Source |
| $As_4Er_2Ni_4SrTb$ | 001 | As | 7.00 | GNoME |
| $DyEu_2Se_3$ | 001 | Se | 6.66 | GNoME |
| $Ba_2MnNbS_6$ | 001 | S | 6.56 | Alexandria |
| $Eu_2PSe$ | 001 | Se | 6.54 | Alexandria |
| $HfNbB_4$ | 001 | B | 6.44 | Alexandria |
| $Sm_2PPb$ | 100 | P | 6.36 | Alexandria |
| $ZrTaB_{32}$ | 001 | B | 6.35 | Alexandria |
| $EuP_2Zr$ | 001 | P | 6.34 | GNoME |
| $EuHfP_2$ | 100 | P | 6.32 | GNoME |
| $TmTaB_{32}$ | 100 | B | 6.32 | Alexandria |

### 3.5 Discussion

Figure 7 illustrates the parity plot between ML-predicted and DFT-calculated work functions for relaxed surfaces in both the low (panel a) and high (panel b) work function regimes. Uncertainty estimates are included for both the ML predictions and the DFT reference values. The ML error bars are the calibrated uncertainty estimates $\hat{\sigma}_{cal}$ described in Sec 3.2, while a fixed uncertainty of ±0.1 eV is assigned to the DFT-calculated work functions based on empirical estimate. For low work functions (Figure 7a), 79% of data points fall within their uncertainty bounds, indicating high prediction confidence and minimal variance. Similarly, for high work functions (Figure 7b), 81% of data points fall "on" the parity line within their uncertainty bounds, suggesting high prediction confidence. The inserted residuals distribution provides a more nuanced view of the prediction errors. For surfaces with low work functions, the distribution of differences between DFT and ML predictions is approximately symmetric around zero, indicating that the ML model neither systematically overestimates nor underestimates. However, for high work function surfaces, the distribution exhibits a noticeable tail toward negative values, suggesting a tendency for the ML model to overpredict work functions.

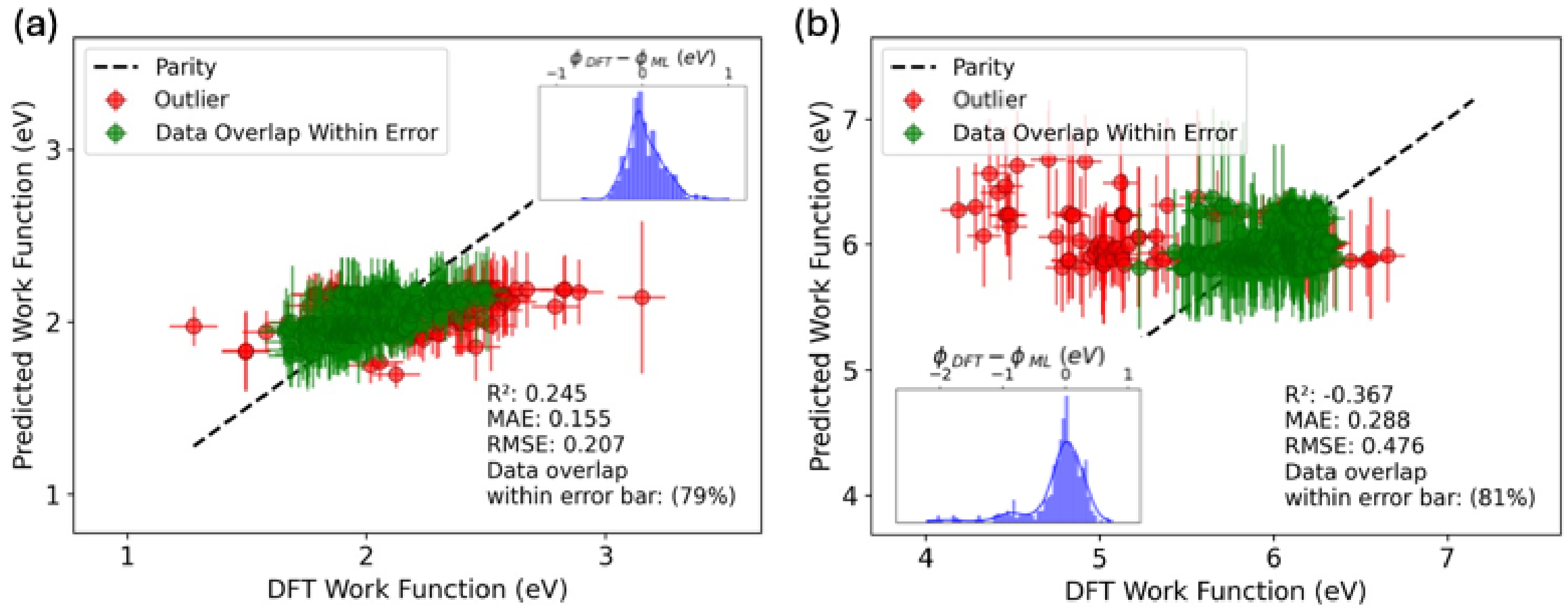

*Figure 7. (a) Parity plot comparing ML-predicted vs. DFT-calculated work functions on relaxed surfaces with low work functions. (b) Parity plot comparing ML-predicted vs. DFT-calculated work functions on relaxed surfaces with high work functions. Inserted pictures illustrate the distribution of residuals.*

To distill chemical insights from these promising candidates, we analyzed the results across various chemical families. Figure 8a illustrates the elemental categories present at the top atomic layers of materials exhibiting extreme work functions. Clear trends emerge from this analysis: surfaces featuring alkali, alkaline earth metals, or lanthanides at the top layer consistently exhibit low work functions. In contrast, surfaces enriched with metalloids, phosphorus, sulfur, or selenium tend to display high work functions. These observations align well with established principles for the rational design of materials with targeted electronic properties, such as that alkali metals, and alkaline earth metals at the top atomic layer tend to exhibit low work functions. Beyond these expected trends, the results also revealed unconventional combinations. Notably, the presence of metalloids or phosphorus at the top layer was found to correlate with exceptionally high work functions, as shown by the two representative surfaces in Figure 8b. Lanthanide-rich surfaces showed a significantly increased likelihood of exhibiting extreme low work functions with two representative surfaces shown in Figure 8b. These findings underscore the value of data-driven screening in uncovering novel chemical configurations that may not be immediately apparent from known design principles.

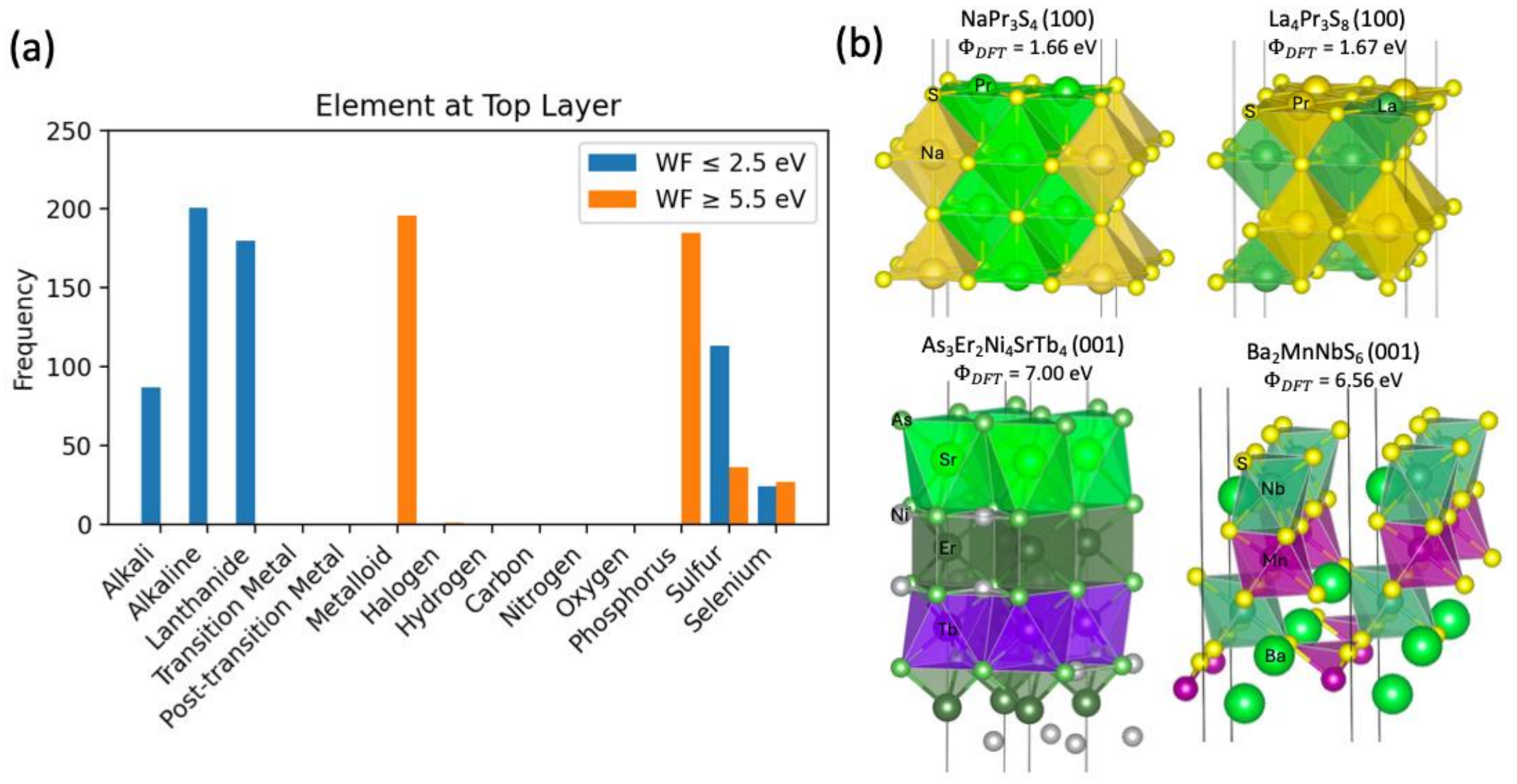

*Figure 8. (a) Frequency distribution of elements present at the top atomic layer of surfaces with extreme work functions. (b) Surface structures of representative candidates with extreme work functions.*

## Summary and Conclusion

In summary, this study presents a robust, multi-fidelity screening approach that integrates machine learning, universal machine learning interatomic potentials M3GNet, and DFT calculations to accelerate the discovery of materials with extreme work functions. We augmented a previously developed random forest model with state-of-the-art uncertainty calibration and domain-of-applicability assessment, enabling more trustworthy predictions for previously unseen materials and surface terminations. We augmented the previously random forest machine learning model with the state-of-the-art uncertainty calibration and domain of applicability assessment, enabling trustworthy predictions on unseen materials and surface terminations. With this augmented model, we screened ≈11 million symmetry-distinct surfaces generated from ≈355k metallic bulk compounds drawn from the GNoME and Alexandria databases, reducing the search space by more than 99.8% before computationally expensive relaxation and DFT calculations. This workflow identified 209 low- ( $\Phi_{DFT} < 2.0$ eV ) and 227 high-work-function ( $\Phi_{DFT} > 6.0$ eV ) surfaces, respectively. Surface relaxation was found to shift predicted values by several hundred meV, demonstrating that incorporating surface relaxation is essential for reliable screening. The chemical intuition extracted from the most promising materials in this study revealed both expected and unconventional trends. The SHAP analysis shows that work function is dominated by electronic descriptors of the surface termination, particularly electronegativity, ionization energy, and atomic size. Surfaces terminated by alkali metals, alkaline earth metals, and lanthanides consistently exhibit low work functions. In contrast, surfaces enriched with metalloids, phosphorus, sulfur, and selenium are statistically associated with high work functions. More broadly, The uncertainty-aware multi-fidelity screening workflow demonstrated here can enable reliable exploration of vast materials spaces while maintaining computational efficiency. The approach highlights the importance of model trustworthiness and physical insight in high-throughput materials discovery and offers practical guidance for future computational and experimental investigations of work function engineering.

**Acknowledgements:** J.M, J.B, and R.J. acknowledge the support of Air Force Office of Scientific Research Grant No. SCON-00004108. R.K acknowledges the support of Air Force Office of Scientific Research Grant FA9550-22-1-0433. This work utilized the computational support from the Advanced Cyberinfrastructure Coordination Ecosystem: Services & Support (ACCESS) program, under Award No. MAT240071, supported by National Science Foundation (NSF) grants #2138259, #2138286, #2138307, #2137603, and #2138296. This research used the computer resources and assistance of the UW – Madison Center for High Throughput Computing (CHTC) in the Department of Computer Sciences.

**Ethical statement:** The authors have no conflicts of interest to declare.

**Data and Code Availability:** All data and codes required to reproduce the findings of this work are available on Figshare: https://doi.org/10.6084/m9.figshare.33201405 and Github: https://github.com/uw-cmg/Uncertainty-Aware-Multi-Fidelity-Screening-.git